\documentclass[12pt]{iopart}

\usepackage{epsfig}
\usepackage{graphicx}
\usepackage{color}
\usepackage[english]{babel}
\usepackage{amssymb}

\newcommand{\bra}[1]{\langle#1|}
\newcommand{\ket}[1]{|#1\rangle}
\newcommand{\beq}{\begin{equation}}
\newcommand{\eeq}{\end{equation}}
\newcommand{\barr}{\begin{eqnarray}}
\newcommand{\earr}{\end{eqnarray}}
\newcommand{\Ham}{\mathcal H}
\newcommand{\eps}{\varepsilon}

\begin{document}
\title{Applicability of the generalized Gibbs ensemble after a quench in the quantum Ising chain}

\author{Tommaso Caneva$^1$, Elena Canovi$^2$, Davide Rossini$^3$, Giuseppe E. Santoro$^{4,5,6}$ and Alessandro Silva$^5$}

\address{$^1$ \, Institut f\"ur Quanten-Informationsverarbeitung, Universit\"at Ulm, D-89069 Ulm, Germany \\
         $^2$ \, Institut f\"ur Theoretische Physik III, Pfaffenwaldring 57, D-70550 Stuttgart, Germany   \\
         $^3$ \, Scuola Normale Superiore, NEST and Istituto Nanoscienze-CNR, I-56126 Pisa, Italy \\
         $^4$ \, SISSA, Via Bonomea 265, I-34136 Trieste, Italy \\
         $^5$ \, International Centre for Theoretical Physics (ICTP), P.O. Box 586, I-34014 Trieste, Italy \\
         $^6$ \, CNR-INFM Democritos National Simulation Center, Via Bonomea 265, I-34136 Trieste, Italy}

\begin{abstract}
  We investigate the out-of-equilibrium dynamics of the one-dimensional quantum Ising model 
  after a sudden quench of the transverse magnetic field.
  While for a translationally invariant system the statistical description of the asymptotic order parameter 
  correlations after the quench can be performed in terms of the generalized Gibbs ensemble, we show that
  a breaking of translational invariance, e.g., by perturbing the boundary conditions, disrupts its validity. 
  This effect, which of course vanishes in the thermodynamic limit, is shown to be very important 
  in the presence of disorder.
\end{abstract}




\section{Introduction}

The experimental advances in the manipulation of ultracold atoms have paved the way
to accurately test the long-time dynamics of quantum many-body systems~\cite{bloch2008}.
Starting from the observation of the collapse and revival of the superfluid order 
parameter after a quench from a shallow to a deep optical lattice~\cite{greiner2002},
a number of beautiful experiments probed the coherent relaxation dynamics 
in interacting quantum systems, with essentially perfect insulation from the external 
environment~\cite{greiner2002,fertig2005,kinoshita06,hofferberth2007,zwierlein2011,bloch11,greiner11}.
Some typical features of adiabatic or quenched dynamics, that are by far harder 
to be identified in actual condensed matter systems, were observed neatly. 
Among the others, they include the phase dynamics emerging after splitting 
a one-dimensional Bose liquid~\cite{hofferberth2007}, or the absence of thermalization 
in out-of-equilibrium arrays of trapped one-dimensional Bose gases~\cite{kinoshita06}.

From a theoretical point of view, this experimental progress spurred renewed interest
on the non-equilibrium dynamics of isolated many-body quantum systems, 
a topic that was previously addressed mostly as an academic question
(for a review, see Ref.~\cite{Polkovnikov2010}). In this context, an issue that received 
a great deal of attention has been the connection between quantum integrability and
thermalization (or its absence) in the dynamics of a closed many-body systems~\cite{Polkovnikov2010}.
The simplest protocol to study it in detail is the so-called {\it quantum quench}, i.e. a sudden 
variation of one of the parameters of the system's Hamiltonian, where the presence or absence of
ergodicity should be detected by studying the state of the system long time 
after the quench has taken place. For non-integrable systems thermalization 
is supposedly occurring at the level of individual eigenstates~\cite{rigolNature} (see also~\cite{Ikeda_preprint10}), 
and ergodic behavior is generally expected to emerge for few body observables~\cite{canovi2011};
actually this scenario seems too broad to be valid in any of such circumstances
and it has been invalidated in some remarkable situations~\cite{biroli2010,banuls2011,Gogolin2011}.
On the other hand, for integrable systems the situation appears to be much less universal: 
non trivial constants of motion generally prevent the system from
thermalizing in the usual sense of standard statistical mechanics.
However, it was argued~\cite{rigolGGE,rigolGGE2} that it is still possible 
to describe the asymptotic state in terms of the generalization of the 
Gibbs ensemble (GGE), first proposed by Jaynes to take into account all constants 
of motion~\cite{jaynesGGE}.

The validity of the GGE for a variety of integrable models has been tested both 
numerically and analytically~\cite{rigolGGE2,cazalilla2006,cardy2007,cramer2008,barthel2008,fioretto2010,cassidy2011}. 
However, since the GGE does not take into account possible correlations 
between different constants of motion in the initial state, it was 
hinted in Ref.~\cite{Gangardt2008} that it should fail as soon as the breaking of global symmetries
(like translational invariance) makes such correlations non negligible. 
The purpose of this paper is to show that this expectation is indeed correct: 
it is enough to slightly perturb the boundary conditions of an integrable model 
(or to make it non-homogeneous) to make the GGE fail.
We will show this on the one-dimensional quantum Ising chain, the simplest non-trivial example 
of an integrable strongly interacting quantum system. 
This model has been the subject of various studies in the 
literature~\cite{McCoy_PRA71,igloi2000,sengupta2004,perk2009,rossini09,rossini09B,fagotti2010,igloi2011,banuls2011,calabrese2011}.
In particular it has been shown semi-numerically that for small quenches the asymptotic behavior 
of the order parameter two point correlation functions displays qualitatively 
a thermal behavior~\cite{rossini09,rossini09B}. On the other hand, as recently shown 
by Calabrese, Essler and Fagotti~\cite{calabrese2011}, the correlation functions 
of the order parameter (as well as of the transverse magnetization) in the translationally 
invariant case can be quantitatively extracted from the GGE, finding agreement 
with the numerical analysis~\cite{rossini09,rossini09B}. Here we will provide evidence that,
if translational invariance is broken, the GGE fails to predict correctly the asymptotic state. 
We will first show this by studying a quench in the quantum Ising chain 
with open boundary conditions: for any finite system we find a significant discrepancy 
between the GGE and the exact analysis, which of course tends to zero in the thermodynamic limit.
These discrepancies become of course very important in the presence of inhomogeneities 
not restricted to the boundaries, as we show by explicitly considering 
the presence of disorder both in the couplings and in the transverse field. 
The Ising model in the presence of completely random couplings and fields has been widely studied, 
see, e.g., Refs.~\cite{Fisher_PRL92,Fisher_PRB95,Young_PRB96,Young_PRB97,Rieger_EPL97,Igloi_PRB98,Caneva_PRB07}; 
here we are more interested in the regime in which the disordered part of the couplings and of the fields
is small, as compared to the uniform part. \par
The paper is organized as follows. In Sec.~\ref{sec:model} we set our notations,
introduce the model under investigation and show the effects of the breaking
of translational invariance on the structure of the eigenmodes of the system.
In Sec.~\ref{sec:quench} we then frame the model in the quantum quench framework,
and define the quantities we are going to discuss subsequently.
Our results are then presented in Sec.~\ref{sec:results}.
Finally we draw our conclusions.

\section{Model} \label{sec:model}

We consider an integrable spin-$1/2$ quantum Ising chain in presence of random on-site
disorder both in the couplings and in the transverse field.
The Hamiltonian for a system of $L$ sites is given by
\beq \label{eq:RImodel}
  \hat{\Ham}_\eps (\Gamma) = - J \sum_{j=1}^{L} (1 + \eps \, \eta_{j}) \hat{\sigma}^{x}_{j} \hat{\sigma}^{x}_{j+1} 
  -\Gamma \sum_{j=1}^{L} (1 + \eps \, h_{j}) \hat{\sigma}^{z}_{j} \,,
\eeq
where $\hat{\sigma}^{\alpha}_{j}$  ($\alpha=x,y,z$) are spin-$1/2$ Pauli matrices for the $i$-the spin, 
$J$ and $\Gamma$ respectively denote the nearest-neighbor coupling and the transverse magnetic
field strength, while $\eta_j, h_j \in [-1,1]$ are dimensionless site-dependent quantities
accounting for any disomogeneity in the model, and $\eps$ sets the strength of the disorder.
In the following we will consider both Open Boundary Conditions (OBC), implemented by setting 
$1 + \eps \, \eta_{L} = 0$ (or equivalently by imposing $\hat{\sigma}^{\alpha}_{L+1} \equiv 0$),  
and Periodic Boundary Conditions (PBC), by assuming $\hat{\sigma}^{\alpha}_{L+1} \equiv \hat{\sigma}^{\alpha}_1$. 
Hereafter we use units in which $\hbar=k_{B}=1$, and set $J=1$ as the energy scale of the system.

The Ising chain can be solved by first performing a Jordan-Wigner 
transformation of the spin-$1/2$ particles into spinless fermions
$\hat{\sigma}^-_j= \exp(-i\pi \sum_{l=1}^{j-1} \hat{c}^{\dagger}_{l} \hat{c}_{l})\hat{c}_{j}$, 
thus mapping Eq.~(\ref{eq:RImodel}) into the following quadratic Hamiltonian:
\barr \label{eq:quadr}
\hat{\Ham}_\eps(\Gamma) & = &
   - \sum_{j=1}^{L-1} (1 + \eps \, \eta_{j}) \left( \hat{c}^{\dagger}_{j} \hat{c}^{\dagger}_{j+1} 
   + \hat{c}^{\dagger}_{j} \hat{c}_{j+1} + H.c. \right)
   - 2 \Gamma \sum_{j=1}^{L} (1 + \eps \, h_{j}) \hat{c}^{\dagger}_{j} \hat{c}_{j}   \nonumber \\
   & & + (-1)^{N_{F}} (1 + \eps \, \eta_{L}) \left(\hat{c}^{\dagger}_{L} \hat{c}^{\dagger}_{1} 
   + \hat{c}^{\dagger}_{L} \hat{c}_{1} + H.c. \right) \, ,
\earr
$N_{F} = \sum_{j=1}^L \hat{c}^{\dagger}_{j} \hat{c}_{j}$ being the total number of $c$-fermions, 
whose parity operator $(-1)^{N_F}$ is conserved, since it commutes with $\hat{\Ham}_\eps(\Gamma)$.
In the general non-uniform case, the system is diagonalized by means of a Bogoliubov 
rotation~\cite{Young_PRB96}, introducing the fermionic operators ($\hat{\gamma}_{\mu}^\dagger, \hat{\gamma}_{\mu}$):
\beq \label{eq:gamma}
   \hat{\gamma}_{\mu} =\sum_{j=1}^{L}(u_{j\mu}^* \hat{c}_{j} + v_{j\mu}^* \hat{c}^{\dagger}_{j}) \,.
\eeq
The $L$-dimensional vectors $\mathbf{u}_{\mu}^T = (u_{1\mu}, u_{2\mu}, \ldots, u_{L\mu})$
and $\mathbf{v}_{\mu}^T = (v_{1\mu}, v_{2\mu}, \ldots, v_{L\mu})$,
for $\mu=1,\dots,L$ can be always chosen to be real, and satisfy the coupled 
Bogoliubov-de Gennes equations:
\beq \label{eq:bog}
\left( \begin{array}{cc} \phantom{-} {\mathbb A} & \phantom{-} {\mathbb B} \\ 
                                  -  {\mathbb B} &          -  {\mathbb A} \end{array}  \right)
\left( \begin{array}{c} \mathbf{u_{\mu}} \\ \mathbf{v}_{\mu}  \end{array} \right)
= \epsilon_{\mu} \left( \begin{array}{c} \mathbf{u}_{\mu}\\ \mathbf{v}_{\mu}  \end{array} \right) \,,
\eeq
where ${\mathbb A}$, ${\mathbb B}$ are $L\times L$ real matrices (${\mathbb A}$ symmetric 
and ${\mathbb B}$ antisymmetric) whose nonzero elements are given by 
${\mathbb A}_{j,j} = -\Gamma (1 + \eps \, h_{j})$, ${\mathbb A}_{j,j+1} = {\mathbb A}_{j+1,j} = -( 1 + \eps \, \eta_{j})/2$, 
${\mathbb B}_{j,j+1} = -{\mathbb B}_{j+1,j} = -( 1 + \eps \, \eta_{j})/2$. 
If PBC are used, then the site $j = L+1$ is identified with $j = 1$ and the following 
matrix elements are also present:
${\mathbb A}_{L,1} =  {\mathbb A}_{1,L} = (-1)^{N_F} (1 + \eps \, \eta_{L})/2$, 
${\mathbb B}_{L,1} = -{\mathbb B}_{1,L} = (-1)^{N_F} (1 + \eps \, \eta_{L})/2$.
In the disordered case $\eps \neq 0$ it is necessary to diagonalize the $2L\times 2L$ 
eigenvalue problem in Eq.~(\ref{eq:bog}) numerically, in order to write
the Hamiltonian $\hat{\Ham}_\eps(\Gamma)$ in terms of its eigenmodes:
\beq \label{diag_H:eq}
\hat{\Ham}_\eps(\Gamma) = 2 \sum_{\mu=1}^{L} \epsilon_{\mu} \left(\hat{\gamma}^{\dagger}_{\mu} \hat{\gamma}_{\mu} 
- \frac{1}{2}\right) \,,
\eeq
where $\epsilon_\mu \geq 0$.
The ground state of $\hat{\Ham}_\eps(\Gamma)$ is the Bogoliubov vacuum state $\ket{\psi_{0}}$
such that $\hat{\gamma}_{\mu} \ket{\psi_{0}} = 0$ ($\forall \mu=1,\ldots, L$), 
with an energy $E_{0} = -\sum_{\mu=1}^{L} \epsilon_{\mu}$.

Before starting the discussion of our problem, let us comment on the form 
of the normal modes for the homogeneous system with PBC and OBC, displaying a crucial difference 
which turns out to be quite important in the quench dynamics.  
In the PBC case the index $\mu$ is the momentum index $k = \pm \pi/L, \, \pm 3\pi/L, \ldots$. 
The system can be equivalently solved by first applying a Fourier transform on the $\hat{c}$-fermions 
and then diagonalizing two-by-two matrices, each one defined in a $(k,-k)$ subspace, a
procedure which is totally independent of $\Gamma$. 
On the contrary, in the OBC case due to the requirement of vanishing wavefunction at the boundaries,
the solution involves standing waves of the form $\sin(kj)$ where, however,
the momenta $k$ obey a quantization condition explicitly involving the transverse field 
parameter $\Gamma$~\cite{Lieb_AP61}:
\begin{equation}
\frac{ \sin{k(L+1)} }{ \sin{kL} } = - \frac{1}{\Gamma} \;.
\end{equation}
As a result, the $k$-vectors are not conserved if the transverse field is changed,
for instance by a quench $\Gamma_0\to \Gamma$ (see below) and the Hilbert space can only 
be split in two subsectors, with odd and even number of fermions respectively,
while $k$-vectors are clearly mixed after changing $\Gamma$. 
A similar situation occurs as soon as $\eps \neq 0$, thus showing 
that the invariant $(k,-k)$ subspace structure for the translational invariant 
situation is very fragile and suddenly disappears in presence of any small imperfection:
It is then unlikely to be the case in experimental implementations of such model.

\section{Quench dynamics and the Generalized Gibbs Ensemble} \label{sec:quench}

The simplest non-equilibrium situation that can be established and controlled
quite feasibly is a quantum quench in the magnetic field strength, that is 
a sudden variation of the value of $\Gamma$ in Eq.~(\ref{eq:RImodel}).
At zero temperature and in absence of any interaction with the environment,
this consists in preparing the system in the ground state $\ket{\psi_{0}}$ 
of $\hat{\Ham}_\eps(\Gamma_0)$ corresponding to a given transverse field $\Gamma_{0}$.
At time $t \equiv 0$, the field is abruptly changed to some different 
value $\Gamma \neq \Gamma_{0}$. 
One then typically follows the unitary evolution under the modified Hamiltonian:
\beq
\ket{\psi (t)} = e^{-i \hat{\Ham}_\eps(\Gamma) t} \ket{\psi_{0}} \, .
\label{eq:psiT}
\eeq
For closed integrable models, the generic absence of thermalization in terms of usual 
statistical ensembles is ascribed to the existence of non trivial integrals of motion.
The way to take into account their effect is to define a Generalized Gibbs Ensemble (GGE)~\cite{rigolGGE},
which explicitly takes into account all such constants of motion $\hat{I}_\mu$ 
through a set of Lagrange multipliers $\lambda_{\mu}$,
\beq
\hat{\rho}_{\rm GGE} = \frac{1}{Z} e^{-\sum_{\mu} \lambda_{\mu} \hat{I}_\mu} \,,
\label{eq:GGE}
\eeq
where $Z={\rm Tr}(\exp[-\sum_{\mu} \lambda_{\mu} \hat{I}_\mu])$. 

In the case of the quantum Ising model, Eq.~(\ref{eq:RImodel}), the number 
$\hat{I}_\mu=\hat{n}_{\mu} \equiv \hat{\gamma}_{\mu}^\dagger \hat{\gamma}_{\mu}$ 
of quasi-particles constitutes a set of integrals of motion: every state can be identified 
in terms of the occupation numbers of all quasi-particle modes.
The average occupation in the initial state can be computed by using the mapping 
from the fermions $\hat{\gamma}^{\Gamma_{0}}_{\mu}$, 
which diagonalize $\hat{\Ham}_\eps(\Gamma_0)$, to the fermions $\hat{\gamma}^{\Gamma}_{\mu}$ 
which diagonalize $\hat{\Ham}_\eps(\Gamma)$:
\barr
 n_{\mu} & = & 
        \bra{\psi_{0}} \hat{\gamma}_{\mu}^{\dagger\Gamma} \hat{\gamma}_{\mu}^{\Gamma} \ket{\psi_{0}} \vspace*{1mm}\\
        & = & \sum_{j l \nu=1}^L 
             \left( v^{\Gamma}_{j\mu} \, u^{\Gamma}_{l\mu} \, u^{\Gamma_{0}}_{j\nu} \, v^{\Gamma_{0}}_{l\nu}
                  + v^{\Gamma}_{j\mu} \, v^{\Gamma}_{l\mu} \, u^{\Gamma_{0}}_{j\nu} \, u^{\Gamma_{0}}_{l\nu}
                  + u^{\Gamma}_{j\mu} \, u^{\Gamma}_{l\mu} \, v^{\Gamma_{0}}_{j\nu} \, v^{\Gamma_{0}}_{l\nu}
                  + u^{\Gamma}_{j\mu} \, v^{\Gamma}_{l\mu} \, v^{\Gamma_{0}}_{j\nu} \, u^{\Gamma_{0}}_{l\nu} \right)
\nonumber 
\earr
One can then construct a GGE starting from these conserved quantities.
The value of each Lagrange multiplier $\lambda_{\mu}$ is fixed by requiring 
$\langle \hat{\gamma}^{\Gamma\dagger}_{\mu} \hat{\gamma}^{\Gamma}_{\mu} \rangle_{\rm GGE} = n_{\mu}$, 
such that $n_{\mu}=\frac{1}{e^{\lambda_{\mu}}+1}$.
Hereafter we denote the expectation value of any observable $\hat{\mathcal O}$ in the GGE with 
$\langle \hat{\mathcal O} \rangle_{\rm GGE} = {\rm Tr}[\hat{\rho}_{\rm GGE} \hat{\mathcal O}]$.

Notice now that for the quantum Ising chain the GGE takes the form of a generalized 
grand-canonical ensemble with a $\mu$ dependent chemical potential. 
In particular, all correlations between the occupations of states with different $\mu$ 
possibly present in the initial state are washed out in the construction. 
While in the ordered PBC case the $(k,-k)$ structure of the solution is preserved
by the quench, so that, for instance
\[ 
\bra{\psi_{0}} \hat{\gamma}^{\dagger\Gamma}_{k'} \hat{\gamma}_{k}^{\Gamma} \ket{\psi_{0}} = 0 \qquad {\rm if} \;\; k'\neq k \;,
\]
with OBC and, more in general, when the system is non homogeneous, the mixing of the eigenmodes leads to 
\[ 
\bra{\psi_{0}} \hat{\gamma}^{\dagger\Gamma}_{\mu} \hat{\gamma}_{\nu}^{\Gamma} \ket{\psi_{0}} \neq 0  \qquad \forall \mu,\nu \;.
\]
As we shall see later, this reflects itself in a breakdown of the validity of the
description of state~(\ref{eq:psiT}) by means of the GGE.

\subsection{Observables}

In the following we will concentrate on the asymptotic behavior of the two-point correlators
of the order parameter $\hat{\sigma}^x$.
We will examine both the equal-time correlator $C^{xx} (0,r)$ of two spins inside the chain 
at distance $r$, and the time-evolved autocorrelation function of the central spin, $C^{xx} (t,0)$.
For the out-of-equilibrium system after the quench, they are explicitly given by
\beq \label{eq:TwoPoint_Q}
C_{Q}^{xx} (t,r) = 
  \bra{\psi (t_0)} \, \hat{\sigma}^{x}_{L/2, [H]} (t)  \, \hat{\sigma}^{x}_{L/2+r} \, \ket{\psi (t_0)} \, ,
\eeq
where $\hat{\mathcal O}_{[H]} (t) \equiv e^{i \hat{\Ham}_\eps (\Gamma) t} \, \hat{\mathcal O} \, e^{-i \hat{\Ham}_\eps (\Gamma) t}$ 
denotes the operator $\hat{\mathcal O}$ in the Heisenberg representation at time $t$,
while $t_0 > 0$ is a lapse time after the quench, that is required for the excited quasiparticles to
propagate along the chain~\footnote{
The lapse time $t_0$ is chosen to be sufficiently large such that $C_{Q}^{xx}(t,r)$ 
at given values of $t,r$ has already reached a steady state, which is independent of $t_0$. 
Conversely, at fixed system size $L$, the lapse time $t_0$ has not to be chosen too large, due
to the appearance of revivals for very long times.
}.

As shown below, for the ordered system ($\epsilon=0$) the correlation functions~(\ref{eq:TwoPoint_Q}) 
decay exponentially both as a function of space and of time: 
$C^{xx}_{Q}(t \gg t_0,r) \sim e^{-r/\xi_{Q}}$ and $C_Q^{xx}(t,0) \sim e^{-t/\tau^{\varphi}_{Q}}$,
the correlation length $\xi_{Q}$ and time $\tau^{\varphi}_{Q}$ are determined,
qualitatively and to a good accuracy even quantitatively, by the canonical ensemble average, that is only through 
an effective temperature set by the initial energy and the final Hamiltonian~\cite{rossini09}.  
The effective temperature, $T_{\rm eff} \equiv \beta_{\rm eff}^{-1}$ is obtained
by comparing the energy of the initial state $\ket{\psi_0}$ with respect to the quenched Hamiltonian
$\hat{\Ham}_\eps(\Gamma)$ to the average energy of a fictitious thermal state in an effective canonical ensemble
$\bra{\psi_0} \hat{\Ham}_\eps(\Gamma_0) \ket{\psi_0} = \langle \hat{\Ham}_\eps(\Gamma) \rangle_{can}$,
where $\langle \hat{\mathcal O} \rangle_{can} = {\rm Tr} (\hat{\mathcal O} e^{-\beta_{\rm eff}\hat{\Ham}_{\eps}})/
{\rm Tr} (e^{-\beta_{\rm eff}\hat{\Ham}_{\eps}})$ denotes the canonical ensemble average~\cite{rossini09}.
Notice that while this effective description in terms of an effective canonical ensemble 
is able to reproduce the gross features of $C^{xx}_{Q}(t,r)$, its fine details require 
a more detailed analysis, which for the translational invariant case (ordered system with PBC)
has been shown to be possible using a GGE~\cite{fagotti2010,calabrese2011}.
In order to assess the degree of validity of GGE in the presence of open boundary conditions, 
and eventually disorder, we will therefore compare the asymptotics of Eq.\ref{eq:TwoPoint_Q} with 
\beq \label{eq:TwoPoint_GGE}
C_{\rm GGE}^{xx}(t,r) = \langle \hat{\sigma}^{x}_{L/2,[H]}(t) \, \hat{\sigma}^{x}_{L/2+r} \rangle_{\rm GGE} \,.
\eeq
Notice that the lapse time $t_0$ has no effects on the GGE, 
since the time evolution operators $e^{-i \hat{\Ham}_\eps(\Gamma) t_0}$ and $e^{i \hat{\Ham}_\eps(\Gamma) t_0}$ 
cancel out inside the trace. 
Since the quantities defined in Eqs.~(\ref{eq:TwoPoint_Q})-(\ref{eq:TwoPoint_GGE}) 
are in general complex, in what follows we will show results for their absolute value. \par
The computation of both the correlators $C_{Q}^{xx}(t,r)$ and $C_{\rm GGE}^{xx}(t,r)$ can be easily implemented numerically, 
because the latter can be written in terms of the fermions ($\hat{\gamma}_{\mu}^\dagger, \hat{\gamma}_{\mu}$) as the
square root of a Pfaffian~\cite{McCoy_PRA71,Lieb_AP61}, see also ~\cite{Young_PRB96,Young_PRB97,Igloi_PRB98}.

\section{Results} \label{sec:results}

\subsection{Different boundary conditions}

The simplest possibility to break the translational invariance
(at least the most harmless, since it corresponds to adding one local defect in
the transverse coupling between two spins, and it becomes irrelevant in the thermodynamic limit)
is to slightly perturb the system by changing its boundary conditions.
We follow this choice and discuss the changes in the asymptotic description of the
order parameter correlator between the homogeneous PBC system and the OBC system.

\begin{figure}[!t]
  \epsfig{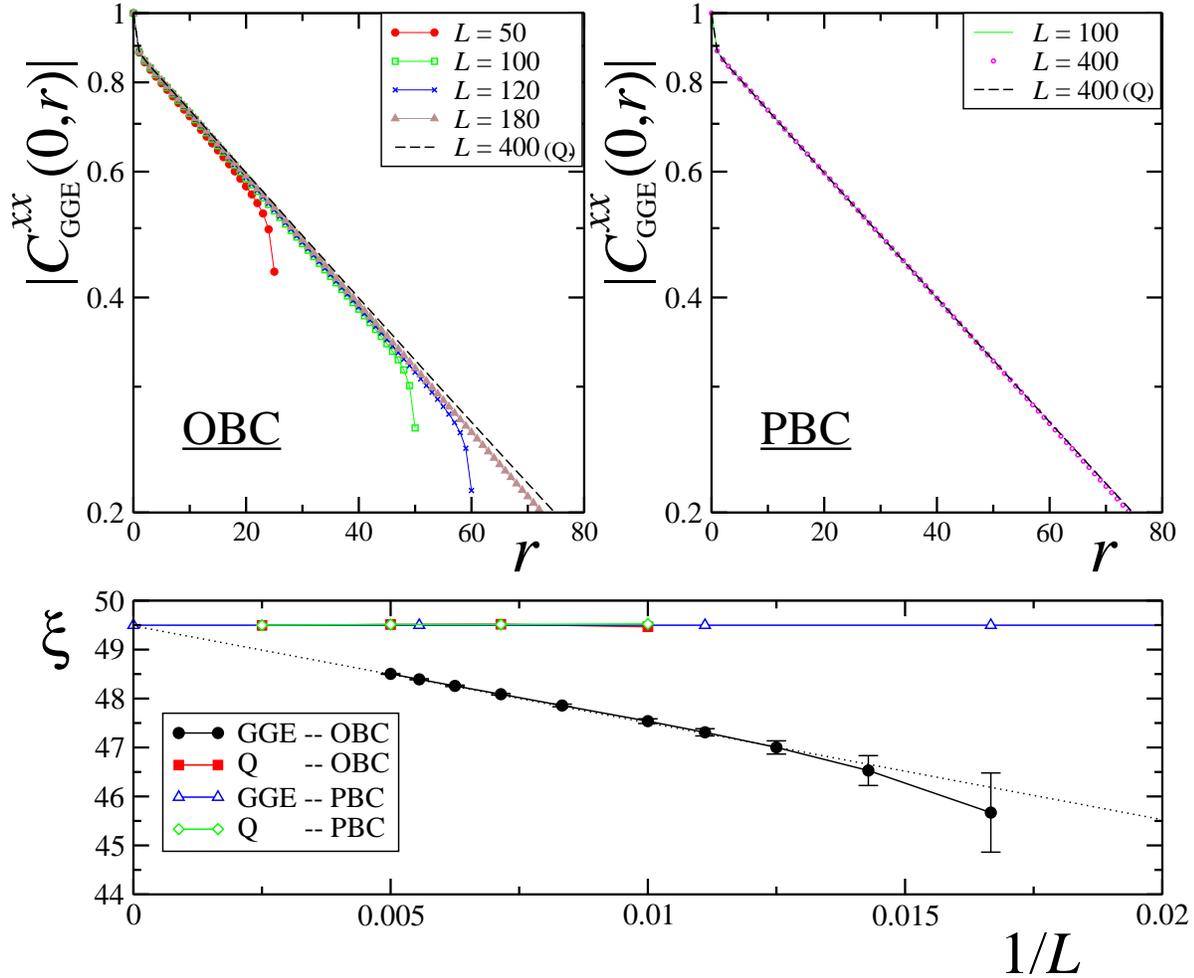}
  \caption{(color online).
    Upper panels: expectation value of the equal-time correlator $\vert C^{xx}(0,r) \vert$ 
    as a function of the distance $r$ from the central site, for OBC (left) and PBC (right),
    after a quench within the Ising ferromagnetic phase from $\Gamma_{0}=0.8$ to $\Gamma=0.6$.
    We show data both for the GGE averages and for the quenched dynamics.
    The colored lines denote GGE averages $C_{\rm GGE}^{xx}(0,r)$ performed 
    at different sizes $L$, as shown in the captions. The black dashed lines 
    indicate $C_Q^{xx}(0,r)$ for $L=400$ sites, and after a time lapse $t_0 = 100$.
    Lower panel: comparison between the correlation length $\xi_{\rm GGE}$ obtained with
    OBC (filled symbols; black circles for the GGE and red squares for the quench)
    and with PBC (empty symbols; blue triangles for the GGE and green diamonds for the quench),
    as a function of the inverse size $1/L$. 
    The dashed black line is a linear fit of the OBC data for the GGE at large sizes.
    Error bars have been estimated by changing the fitting interval over the data for $r < L/2$.}
  \label{fig:sOBC}
\end{figure}

Let us begin with the spatial decay of the two-point correlators $\vert C^{xx}(0,r) \vert$,
and study the discrepancies between the quenched dynamics,
$\vert C^{xx}_{Q}(0,r) \vert$, and the GGE averages $\vert C^{xx}_{\rm GGE}(0,r) \vert$. 
As it is apparent from the upper panels of Fig.~\ref{fig:sOBC}, 
the correlation functions always decay exponentially with the distance in both cases,
and both for OBC and PBC: $\vert C^{xx}(0,r) \vert \sim e^{-t/\xi}$, 
where $\xi$ denotes the correlation length.
However, while for PBC there is substantial agreement between the asymptotic correlator 
and that predicted by the GGE (as it is apparent from the right panel), this agreement 
is definitely disrupted for OBC (left panel). In this case, contrary to the quenched case,
the GGE data exhibit a dependence on the size $L$; this introduces some discrepancies
with the out-of-equilibrium case (dashed line).
The size-dependence is highlighted in the lower panel, where we plotted
the correlation length $\xi_Q$ and $\xi_{\rm GGE}$ as a function of the inverse size $1/L$.
Summarizing, the values of $\xi_Q$ after a quench are essentially independent of the size 
and the boundary conditions (for the case presented here with $\Gamma_{0}=0.8$ and $\Gamma=0.6$, 
for example, we found $\xi_Q \approx 49.5$) and agree well with $\xi_{\rm GGE,\, PBC}$; 
vice-versa, $\xi_{\rm GGE,\, OBC}$ displays a systematic discrepancy which apparently 
vanishes in the thermodynamic limit $L\to \infty$.
Ultimately, the discrepancies found between GGE and quenches with OBC are due to the fact that 
the initial state of the quench 
and the Hamiltonian after the quench are governed by different sets of $k$-values,
as remarked in Sec.~\ref{sec:model}.

\begin{figure}[!t]
  \epsfig{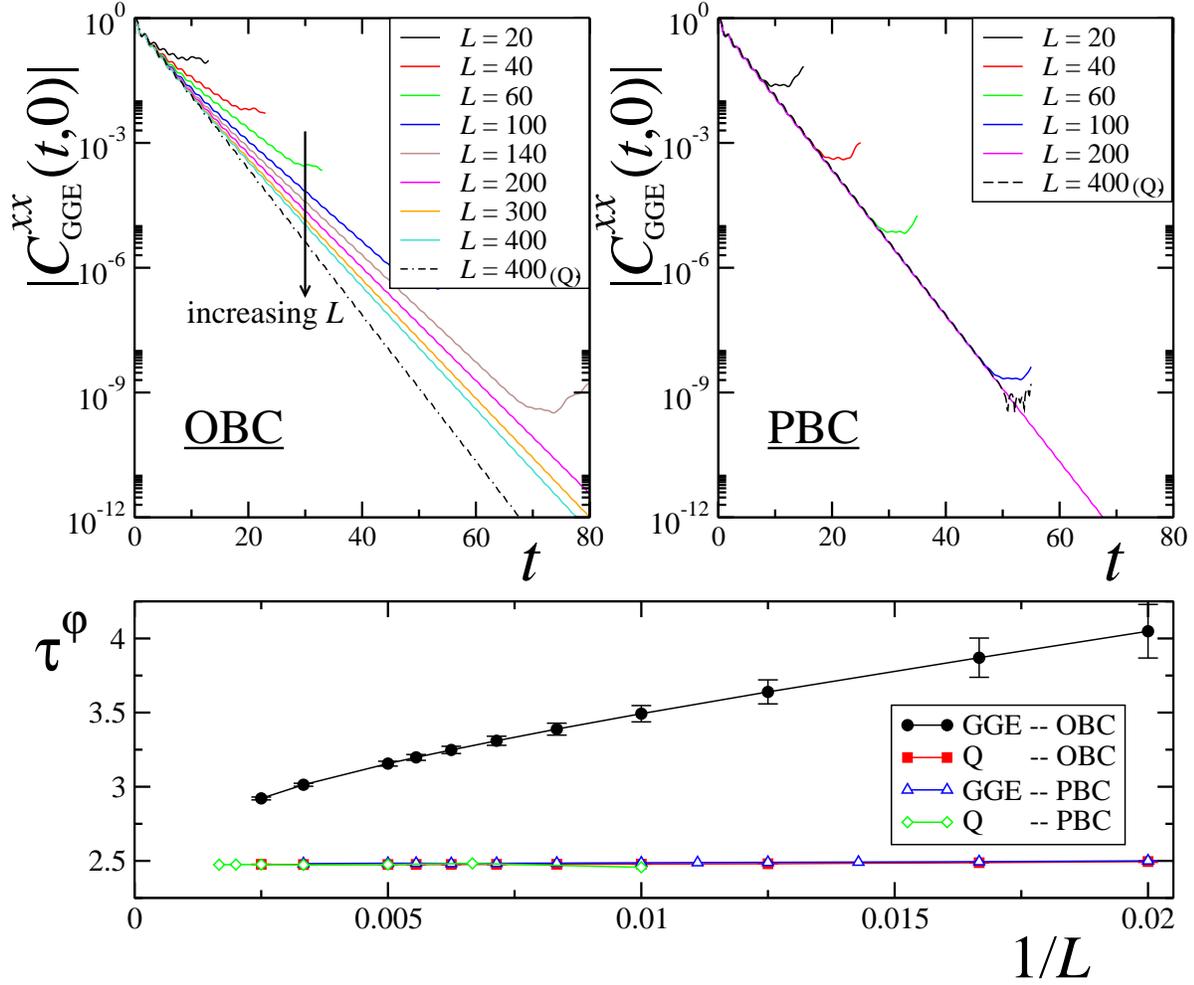}
  \caption{(color online). 
    Upper panels: expectation values of the autocorrelation function $\vert C^{xx}(t,0) \vert$ 
    on the central site as a function of time $t$, computed for different sizes $L$, with OBC (left)
    and PBC (right). 
    All the curves denote GGE averages, except the dashed black line which is for the quenched dynamics.
    A quench from the Ising-paramagnetic phase $\Gamma_{0}=1.6$ to the critical point $\Gamma=1$ has been performed. 
    Lower plot: comparison between the decay rate $\tau_{Q}^{\varphi}$ and $\tau_{\rm GGE}^{\varphi}$ 
    both for OBC and PBC, as functions of the inverse size $1/L$.
    An estimate of the error bars is obtained similarly to Fig.~\ref{fig:sOBC}.}
  \label{fig:tOBC}
\end{figure}

We now turn to the on-site correlator $\vert C^{xx}(t,0) \vert$ as a function of time. 
As for the spatial decay, also in this case we observe an exponential decay:
$\vert C^{xx}(t,0) \vert \sim e^{-t/\tau^{\varphi}}$, until a time $t^*$ 
where finite-size revivals appear~\footnote{
A rough estimate of the revival times is given by $t^* = N/\bar{v}_\Gamma$,
where $\bar{v}_\Gamma$ is the maximum phase velocity of the spin chain over
all the normal-mode velocities. Given the dispersion relation $E_k$, then 
$\bar{v}_\Gamma = {\rm Max}_{\bar{k}} (\partial_k E_k) \vert_{\bar{k}}$.
For the uniform Ising chain with PBC, the maximum can be easily 
evaluated to give $\bar{v}_\Gamma = 2 \, {\rm Min} (\Gamma,1)$.}. 
This occurs independently of the choice of the boundary conditions, and both for the system after the
quench and for the GGE.
The differences in the GGE average between OBC and PBC emerge clearly also here,
and closely resemble what we already observed for the spatial correlators.
In the first case (upper left panel), the size dependence of $\vert C^{xx}_{\rm GGE}(t,0) \vert$
is evident. On the contrary, the PBC-case is basically insensitive to $L$ (upper right panel)~\footnote{
The computation of the autocorrelation function after a quench with PBC is not straightforward. 
The difficulty lays in the fact that $\sigma^x_{j,[H]}(t) \sigma^x_j$ connects states 
with different $c$-fermion parity which are therefore subject to different 
boundary conditions in the PBC case, see the last term in the r.h.s. of Eq.~(\ref{eq:quadr}).
We overcame this point by evaluating a $c$-parity conserving 
four-fermion correlator and then took its square root~\cite{McCoy_PRA71}.
}.
The out-of-equilibrium correlation function $C^{xx}_Q(t,0)$ after a quench 
appears to be essentially independent of the size and of the boundary conditions, 
for times shorter than the revival time $t^*$.
If we compare the decay rate $\tau^{\varphi}_Q$ (in the case highlighted in the bottom panel 
of Fig.~\ref{fig:tOBC}, where we presented data for a quench from $\Gamma_{0}=1.6$ to $\Gamma=1$,
we found $\tau^{\varphi}_Q \approx 2.48$) with the rate  $\tau^{\varphi}_{\rm GGE}$ extracted 
by averaging over the GGE, we found a quantitative agreement only for PBC, 
while some discrepancies are present for the decay rate averaged with open boundaries.
In particular, at a given finite size, $\tau_{\rm Q, \, OBC}^{\varphi} \neq \tau_{\rm GGE,\, OBC}^{\varphi}$, 
even if the agreement improves for increasing sizes. 
Despite the fact that we cannot be conclusive with only numerical results,
for $L\to\infty$ the results appear to have a tendency towards a convergence to a unique decay rate
equal to the one for the quenched dynamics in the thermodynamic limit, where 
boundary conditions should be irrelevant.

\subsection{Effect of small disorder}

There are many other possible ways of breaking translational invariance.
Disorder in the system destroys such invariance everywhere, in a much
more substantial way than by changing the boundary conditions.
Here we discuss what happens if a random component in the couplings and the magnetic fields is added,
$\eps \neq 0$ in Eq.~(\ref{eq:RImodel}).

\begin{figure}[!t]
  \epsfig{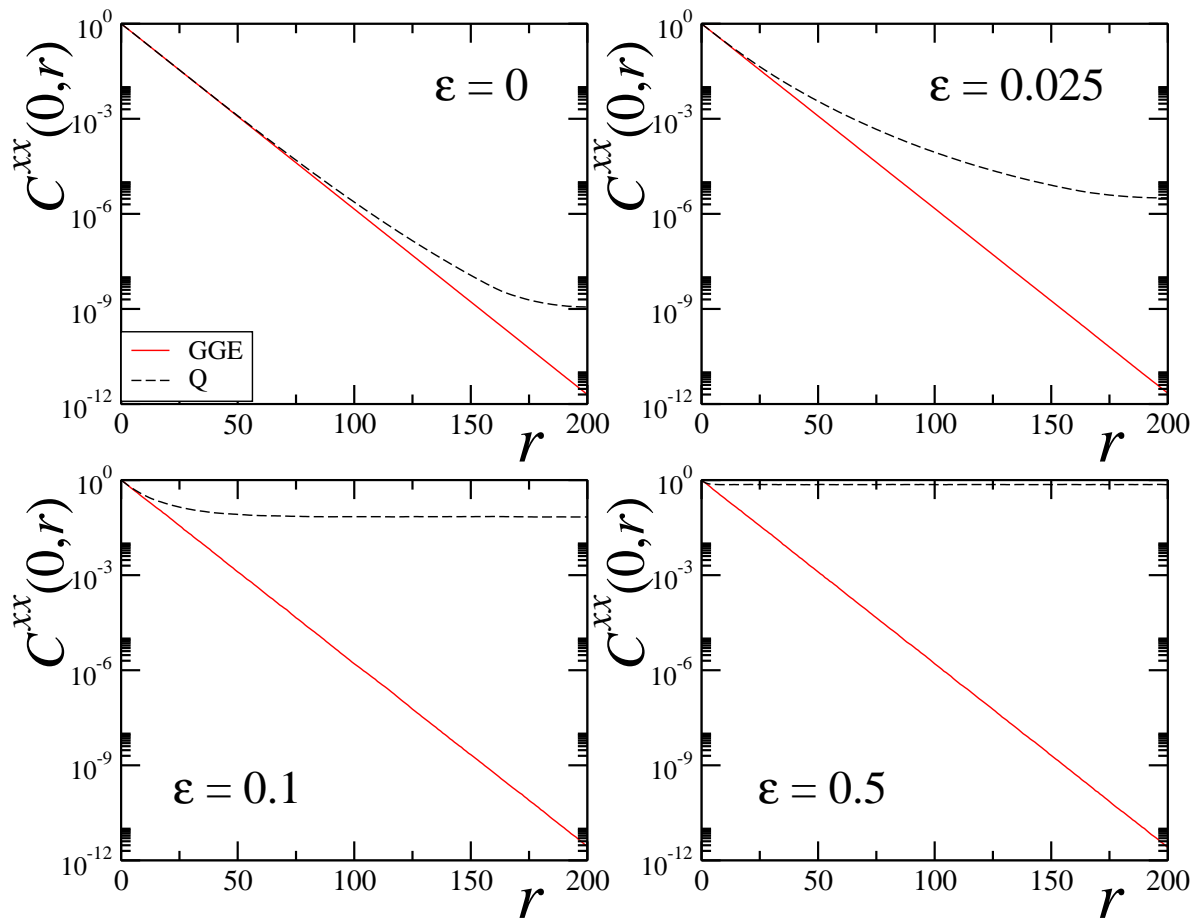}
  \caption{(color online). 
    Spatial decay of the equal-time correlator $\vert C^{xx}(0,r) \vert$ 
    in the disordered Ising model. The various panels are for different values of the
    disorder strength $\eps = $ 0 (upper left), 0.025 (upper right), 0.1 (lower left), 0.5 (lower right).
    The dashed black lines are for the out-of-equilibrium system after a quench 
    from $\Gamma_0 =0.8$ to $\Gamma = 0.2$ (here we set a lapse time $t_0 = 300$), 
    while the continuous red curves denote the corresponding GGE averages.
    A system of $L = 400$ sites has been considered. 
    We performed averages over 200 and 25 disorder realizations, respectively, 
    for the quenched dynamics and for the GGE average.}
  \label{fig:dPBC}
\end{figure}

As it is apparent from Fig.~\ref{fig:dPBC}, where we analyzed the spatial decay of the
two-point correlator of the order parameter, the differences between the quenched dynamics
and the GGE averages are striking. 
First of all we note that $\vert C^{xx}_Q(0,r) \vert$ tends to a constant value for $r \to \infty$,
while  $\vert C^{xx}_{\rm GGE} (0,r) \vert$ decays exponentially, irrespective of the value of $\eps$
(remarkably, the effect of the noise in the GGE average is basically negligible, on the scale
of the figures).
This clearly shows that the validity of the GGE suddenly ceases, as soon as a tiny
amount of disorder is added in the system.
In the panels of Fig.~\ref{fig:dPBC}, we observe a quantitative agreement of the GGE
average only in the first panel which corresponds to the limiting case of zero disorder.
The differences between the two curves for $r \gtrsim 100$ are due to finite-size
effects.
We remark that, in order to minimize fluctuations due to the disorder, we performed averages
over different noise realizations. We found that, while for the quenched dynamics this was
essential, the GGE averages are basically unaffected by the disorder: 
already with a single realization of the couplings a clean exponential decay clearly emerges 
(see the red curves in the figure).

\section{Conclusions} \label{sec:concl}

We explicitly tested the predictions of the generalized Gibbs ensemble 
in the 1D quantum Ising model.
We have shown that the GGE is able to provide a quantitative description 
of the asymptotic behavior of the order parameter correlation function
only in the case in which there are no correlations between the different constants of motion.
This corresponds to the homogeneous, translationally invariant, situation~\cite{calabrese2011}.
On the other hand, as soon as the translational invariance is broken,
the GGE fails to apply. 

In particular, we performed two different analyses.
First, we added a local perturbation, by changing the boundary conditions from PBC to OBC:
in this case, even though the GGE still exhibits a qualitative agreement with
the out-of-equilibrium dynamics, a quantitative prediction is recovered only in the thermodynamic limit
$L\to \infty$.
Second, we perturbed the system globally, 
by adding disorder both in the interaction coupling strengths and in the transverse field:
in this case, the effects of the translational invariance breaking are much more evident:
even a very weak disorder leads to a complete failure of the GGE in describing the quench dynamics.
We found that, for the dynamics after the quench in presence of disorder, 
it is no longer possible to define a typical relaxation time, 
in accordance with what recently noticed in Ref.~\cite{Garcia_preprint11}.
This is due to the localization effects, playing here a major role.
Quite surprisingly, the GGE predictions are barely affected by such effects, thus
exhibiting a qualitatively different (exponential) decay of the correlation functions. 

\section*{Acknowledgments}
We thank P. Calabrese, R. Fazio, D. Fioretto and S. Ziraldo for useful discussions. 
T.C. acknowledges AQUTE for the financial support and the BW-grid (http://www.bw-grid.de) for the computational resources.
E.C. acknowledges SISSA, SNS for hospitality and eLab (http://www.escience-lab.org/) for the computational resources. 
D.R. acknowledges support from EU through the project SOLID and from
the FIRB IDEAS project RBID08B3FM.
G.E.S. acknowledges partial support from the Italian Ministry of University and Research, 
through a PRIN/COFIN contract.



\section*{References}

\begin{thebibliography}{99}

\bibitem{bloch2008}
  Bloch I, Dalibard J and Zwerger W
  2008 {\it Rev. Mod. Phys.} {\bf 80} 885

\bibitem{greiner2002}
  Greiner M, Mandel O, H\"ansch T and Bloch I
  2002 {\it Nature} {\bf 419} 51

\bibitem{fertig2005}
  Fertig C D, O’Hara K M, Huckans J H, Rolston S L, Phillips W D and Porto J V 
  2005 {\it Phys. Rev. Lett.} {\bf 94} 120403

\bibitem{kinoshita06}
  Kinoshita T, Wenger T and Weiss D S
  2006 {\it Nature} {\bf 440} 900

\bibitem{hofferberth2007}
  Hofferberth S, Lesanovsky I, Fischer B, Schumm T and Schmiedmayer J
  2007 {\it Nature} {\bf 449} 324

\bibitem{zwierlein2011}
  Sommer A, Ku M, Roati G and Zwierlein M W
  2011 {\it Nature} {\bf 472} 201

\bibitem{bloch11}
  Trotsky S, Chen Y-A, Flesch A, McCulloch I P, Schollw\"ock U, Eisert J and Bloch I
  arXiv:1101.2659

\bibitem{greiner11}
  Simon J, Bakr W S, Ma R, Tai M E, Preiss P M and Greiner M
  2011 {\it Nature} {\bf 472} 307

\bibitem{Polkovnikov2010} 
  Polkovnikov A, Sengupta K, Silva A and Vengalattore M
  arXiv:1007:5331

\bibitem{rigolNature}
  Rigol M, Dunjko V and Olshanii M
  2008 {\it Nature} {\bf 452} 854

\bibitem{Ikeda_preprint10}
  Ikeda T N, Watanabe Y and Ueda M
  arXiv:1012.3237

\bibitem{canovi2011}
  Canovi E, Rossini D, Fazio R, Santoro G E and Silva A 
  2011 {\it Phys. Rev. B} {\bf 83} 094431 

\bibitem{biroli2010}
  Biroli G, Kollath C and L\"auchli A M
  2010 {\it Phys. Rev. Lett.} {\bf 105} 250401

\bibitem{banuls2011}
  Ba\~nuls M C, Cirac J I and Hastings M B
  2011 {\it Phys. Rev. Lett.} {\bf 106} 050405

\bibitem{Gogolin2011}
  Gogolin C, M\"uller M P and Eisert J
  2011 {\it Phys. Rev. Lett.} {\bf 106} 040401

\bibitem{rigolGGE}
  Rigol M, Dunjko V, Yurovsky V and Olshanii M
  2007 {\it Phys. Rev. Lett.} {\bf 98} 050405

\bibitem{rigolGGE2}
  Rigol M, Muramatsu A and Olshanii M
  2006 {\it Phys. Rev. A} {\bf 74} 053616

\bibitem{jaynesGGE}
  Jaynes E T
  1957 {\it Phys. Rev.} {\bf 106} 620;
  1957 {\it Phys. Rev.} {\bf 108} 171

\bibitem{cazalilla2006}
  Cazalilla M A
  2006 {\it Phys. Rev. Lett.} {\bf 97} 156403

\bibitem{cardy2007}
  Calabrese P and Cardy J
  2007 {\it J. Stat. Mech.} P06008

\bibitem{cramer2008}
  Cramer M, Dawson C M, Eisert J and Osborne T J
  2008 {\it Phys. Rev. Lett.} {\bf 100} 030602

\bibitem{barthel2008}
  Barthel T and Schollw\"ock U
  2008 {\it Phys. Rev. Lett.} {\bf 100} 100601

\bibitem{fioretto2010}
  Fioretto D and Mussardo G
  2010 {\it New J. Phys.} {\bf 12} 055015

\bibitem{cassidy2011}
  Cassidy A C, Clark C W and Rigol M
  2011 {\it Phys. Rev. Lett.} {\bf 106} 140405

\bibitem{Gangardt2008}
  Gangardt D M and Pustilnik M
  2008 {\it Phys. Rev. A} {\bf 77} 041604(R)

\bibitem{McCoy_PRA71}
  McCoy B M, Barouch E and Abraham D B
  1971 {\it Phys. Rev. A} {\bf 4} 2331

\bibitem{igloi2000}
  Igl\'oi F and Rieger H
  2000 {\it Phys. Rev. Lett.} {\bf 85} 3233

\bibitem{sengupta2004}
  Sengupta K, Powell S and Sachdev S
  2004 {\it Phys. Rev. A} {\bf 69} 053616

\bibitem{perk2009}
  Perk J H H and Au-Yang H
  2009 {\it J. Stat. Phys.} {\bf 135} 599

\bibitem{rossini09}
  Rossini D, Silva A, Mussardo G and Santoro G E
  2009 {\it Phys. Rev. Lett.} {\bf 102} 127204

\bibitem{rossini09B}
  Rossini D, Suzuki S, Silva A, Mussardo G and Santoro G E
  2010 {\it Phys. Rev. B} {\bf 82} 144302

\bibitem{fagotti2010}
  Fagotti M and Calabrese P
  2008 {\it Phys. Rev. A} {\bf 78} 010306(R)

\bibitem{igloi2011}
  Igl\'oi F and Rieger H
  2011 {\it Phys. Rev. Lett.} {\bf 106} 035701

\bibitem{calabrese2011}
  Calabrese P, Essler F H L and Fagotti M
  2011 {\it Phys. Rev. Lett.} {\bf 106} 227203

\bibitem{Fisher_PRL92}
  Fisher D S
  1992 {\it Phys. Rev. Lett.} {\bf 69} 534

\bibitem{Fisher_PRB95}
  Fisher D S
  1995 {\it Phys. Rev. B} {\bf 51} 6411

\bibitem{Young_PRB96}
  Young A P and Rieger H
  1996 {\it Phys. Rev. B} {\bf 53} 8486

\bibitem{Young_PRB97}
  Young A P
  1997 {\it Phys. Rev. B} {\bf 56} 11691
 
\bibitem{Rieger_EPL97}
  Rieger H and Igl\'oi F
  1997 {\it Europhys. Lett.} {\bf 39} 135

\bibitem{Igloi_PRB98}
  Igl\'oi F and Rieger H
  1998 {\it Phys. Rev. B} {\bf 57} 11404
  
\bibitem{Caneva_PRB07}
  Caneva T, Fazio R and Santoro G E
  2007 {\it Phys. Rev. B} {\bf 76} 144427

\bibitem{Lieb_AP61}
  Lieb E, Schultz T and Mattis D
  1961 {\it Ann. Phys.} {\bf 16} 407

\bibitem{Garcia_preprint11}
  Garc\'ia-Garc\'ia A M, Rela\~no A and Rigol M
  arXiv:1103.0787

\end{thebibliography}
\end{document}